\documentclass{ws-ijqi}
\usepackage{amssymb}

\newcounter{myctr}
\def\myitem{\refstepcounter{myctr}\bibfont\noindent\ifnum\themyctr>9\else\phantom{0}\fi\hangindent17pt\themyctr.\enskip}

\input{tcilatex}
\begin{document}

\title{QUANTUM-STATE TRANSFER BETWEEN ATOM AND MACROSCOPICALLY
DISTINGUISHABLE CAVITY FIELD IN JAYNES-CUMMINGS MODEL}
\author{GANG ZHANG \and GUAN-RONG LI \and ZHI SONG}
\maketitle

\begin{abstract}
We present a scheme for transferring quantum state between atom and cavity
field in Jaynes-Cummings model in the aid of spin-echo-like technique. It is
based on the facts that the atom in a cavity can induce the generation of
modified coherent states, which can be shown to be macroscopically
distinguishable, and the anti-commutation relation between the Hamiltonian
and the $z$-component Pauli matrix. We show that this scheme is applicable
for a class of cavity field states. The application on two-cavity system
provides an alternative scheme for preparation of non-local superpositions
of quasi-classical light states. Numerical simulation shows that the
proposed schemes are efficient.
\end{abstract}


\markboth{Gang Zhang, Guan-rong Li, Zhi Song}
{Quantum-state transfer between atom and macroscopically distinguishable cavity
field in Jaynes-Cummings model}

\catchline{}{}{}{}{}

\address{School of Physics, Nankai University, Tianjin 300071, China}

\address{School of Physics, Nankai University, Tianjin 300071, China}

\address{School of Physics, Nankai University, Tianjin 300071, China}

\begin{history}
\received{Day Month Year}
\revised{Day Month Year}
\end{history}




\section{Introduction}

Coherent transfer between an arbitrary state of a qubit and a superposition
of two quasi-classical coherent states is of fundamental conceptual interest
in many fields of physics \cite%
{JAWheeler83,CCGerry05,STakagi02,MANielsen01,CEAJarvis09}. Coherent states
provide a close connection between classical and quantum mechanics, which
has been introduced in a physical context, first as quasi-classical states
in quantum mechanics, then as the backbone of quantum optics \cite%
{J-PGazeau09}. The non-classical nature of such states appears since two
coherent states correspond to two different values of a macroscopic
variable, such as the quasi-probability distribution in phase space.
Recently, in a new branch of quantum computing, two phase-opposite coherent
states are exploited to be the macroscopical qubits\cite%
{PCochrane99,HJeong02,TCRalph03,TCRalph08}. Much attention has been paid on
obtaining such superposed coherent states \cite%
{AOurjoumtsev07,JSNeergaard-Nielsen06,HTakahashi08,TGerrits11}.

In this paper, we propose a type of modified coherent states based on the
canonical coherent state, which also demonstrate the non-classical nature.
We show that an arbitrary atom state can be mapped onto the field as a Schr%
\"{o}dinger cat state in a resonant Jaynes-Cummings (JC) model. The reversal
process can also be achieved in the aid of spin-echo-like technique. It is
shown that a class of field states can be used as an initial state to
perform this task.\ Numerical simulation shows that the scheme is efficient
and can be extended to the entanglement transfer from atoms to distant
cavities. This make it possible to realize entangled pairs of macroscopic
objects, nonlocal Schr\"{o}dinger cat state.

This paper is organized as follows. In Section 1, we present a modified
coherent state which is shown to be equivalent to a canonical coherent
state. In Section 2, we investigate the JC model and propose the effective
Hamiltonian for a type of states. In Section 3, we study the dynamics of the
system for the initial field state being a coherent state. In Section 4, as
an application we investigate the entanglement transfer between atoms and
fields in the two-cavity system. Finally, we give a summary in Section 5.

\section{Modified coherent state}

\label{Sec_Modified coherent state}We start with a canonical coherent state
\begin{equation}
\left\vert \alpha \right\rangle =\mathrm{e}^{-\left\vert \alpha \right\vert
^{2}/2}\sum_{n}\frac{\alpha ^{n}}{\sqrt{n!}}\left\vert n\right\rangle ,
\label{CS}
\end{equation}%
which is the eigen state of the boson annihilation operator $a$, and the
Fock state is defined as $\left\vert n\right\rangle =\left( a^{\dag }\right)
^{n}/\sqrt{n!}\left\vert \mathrm{Vac}\right\rangle $\ with $\left\vert
\mathrm{Vac}\right\rangle $\ being the vacuum state of $a$. It is a Gaussian
wavepacket in the coordinate representations $x$ $=1/\sqrt{2}\left( a^{\dag
}+a\right) $, whose center is shifted $\sqrt{2}\mathrm{Re}(\alpha )$\ from
the origin. The amplitude $\left\vert \alpha \right\vert $\ characterizes
the distance between $\left\vert \alpha \right\rangle $\ and $\left\vert
-\alpha \right\rangle $\ in phase space. Then the states $\left\vert \alpha
\right\rangle $\ and $\left\vert -\alpha \right\rangle $\ are sufficiently
distinguishable for large $\left\vert \alpha \right\vert $, and present many
advantages compared with discrete variable qubit states $\left\vert
0\right\rangle $\ and $\left\vert 1\right\rangle $.

Now we consider a type of modified coherent state (MCS)
\begin{equation}
\left\vert \alpha ,g\right\rangle =\mathrm{e}^{-\left\vert \alpha
\right\vert ^{2}/2}\sum_{n}\frac{\alpha ^{n}}{\sqrt{n!}}\mathrm{e}^{-\mathrm{%
i}g\left( n\right) }\left\vert n\right\rangle ,  \label{MCS}
\end{equation}%
where $g\left( n\right) $ is an arbitrary real function. Here the term
modified\textit{\ }just indicates the difference from the canonical coherent
state $\left\vert \alpha \right\rangle $, which is the simplest case of the
MCS with $g\left( n\right) =0$, \textit{i.e.}, $\left\vert \alpha
\right\rangle =\left\vert \alpha ,0\right\rangle $. It is worth nothing
that\ the MCS $\left\vert \alpha ,g\right\rangle $\ is equivalent to the
state $\left\vert \alpha ,0\right\rangle $\ when associated with a
Hamiltonian in the form of $h=h(a^{\dag }a)$, where $h$\ is an arbitrary
function.

Based on the original boson operator $a$,\ let us now define a class of
boson operator%
\begin{equation}
b=\mathrm{e}^{\mathrm{i}\left[ g\left( a^{\dag }a\right) -g\left( a^{\dag
}a+1\right) \right] }a,  \label{b}
\end{equation}%
with an arbitrary real function $g$ as long as $g\left( 0\right) =0$. It
turns out that $b$\ satisfies the commutation relation $\left[ b,b^{\dag }%
\right] =1.$ Then the Fock state associated with the annihilation operator $%
b $\ can be written as $\left\vert n,g\right\rangle =\frac{\left( b^{\dag
}\right) ^{n}}{\sqrt{n!}}\left\vert \mathit{Vac}\right\rangle ,$ based on
that the modified coherent state $\left\vert \alpha ,g\right\rangle $\ has
the standard form $\left\vert \alpha ,g\right\rangle =\mathrm{e}%
^{-\left\vert \alpha \right\vert ^{2}/2}\sum_{n}\frac{\alpha ^{n}}{\sqrt{n!}}%
\left\vert n,g\right\rangle .$ We see that the MCSs $\left\vert \alpha
,g\right\rangle $\ and $\left\vert \alpha ,0\right\rangle $\ are connected
by the transformation in Eq. (\ref{b}). We would like to point that, there
is no essential connection between $g\left( n\right) $\ and $h(n)$. However,
$h(a^{\dag }a)$\ can be transformed to a form of $h(b^{\dag }b)$, via the
transformation in Eq. (\ref{b}). In other words, $h(a^{\dag }a)$\ is
covariant under this transformation. In this sense, bosons $a$\ and $b$, as
well as coherent states $\left\vert \alpha ,0\right\rangle $\ and $%
\left\vert \alpha ,g\right\rangle $\ are absolutely\ equivalent for a system
described by $h$. Hereafter we will not distinguish between the standard
and\ the modified coherent states.

In this paper, we focus on a simple case with $g\left( n\right) $ $=\gamma
\sqrt{n},$ which generates the coherent state%
\begin{equation}
\left\vert \alpha ,\gamma \right\rangle =\mathrm{e}^{-\left\vert \alpha
\right\vert ^{2}/2}\sum_{n}\frac{\alpha ^{n}}{\sqrt{n!}}\mathrm{e}^{-\mathrm{%
i}\gamma \sqrt{n}}\left\vert n\right\rangle .
\end{equation}%
According to the above analysis, coherent states $\left\vert \pm \alpha
,0\right\rangle $\ are sufficiently distinguishable for large $\left\vert
\alpha \right\vert $. In this paper, we are interested in the pair of states
$\left\vert \alpha ,\pm \gamma \right\rangle $. We will show that the
atom-field coupling can induce the generation of state $\left\vert \alpha
,\pm \gamma \right\rangle $ from $\left\vert \alpha ,0\right\rangle $ by
natural time evolution and two states $\left\vert \alpha ,\pm \gamma
\right\rangle $\ are macroscopically distinguishable as that of $\left\vert
\pm \alpha ,0\right\rangle $. To this end, we compute the Wigner
quasiprobability distribution $W_{\alpha ,\gamma }(x,p)$\ in phase space,
where\ $x=\left( a+a^{\dag }\right) /\sqrt{2}$, $p=\left( a-a^{\dag }\right)
/\left( \mathrm{i}\sqrt{2}\right) $\ are the quadrature operators of the
cavity field. In quantum optics the Wigner quasiprobability distribution
play an important role. With the help of the Wigner quasiprobability
distribution, we can know which states are differentiable. Up until now, the
different schemes proposed so far to determine the Wigner distribution of a
quantum system rely either on tomographic reconstructions from data obtained
in homodyne measurements or on convolutions obtained by photon counting\cite%
{Lutterbach96,Vogel89,Smithey93,Leonhardt94,Ariano95,Dunn95,Wallentowitz95,Helon96,Poyatos96,Banaszek96,Wallentowitz96}%
\textbf{.}

\begin{figure}[tbp]
\begin{center}
\includegraphics[bb=0 0 600 450, width=0.45\textwidth, clip]{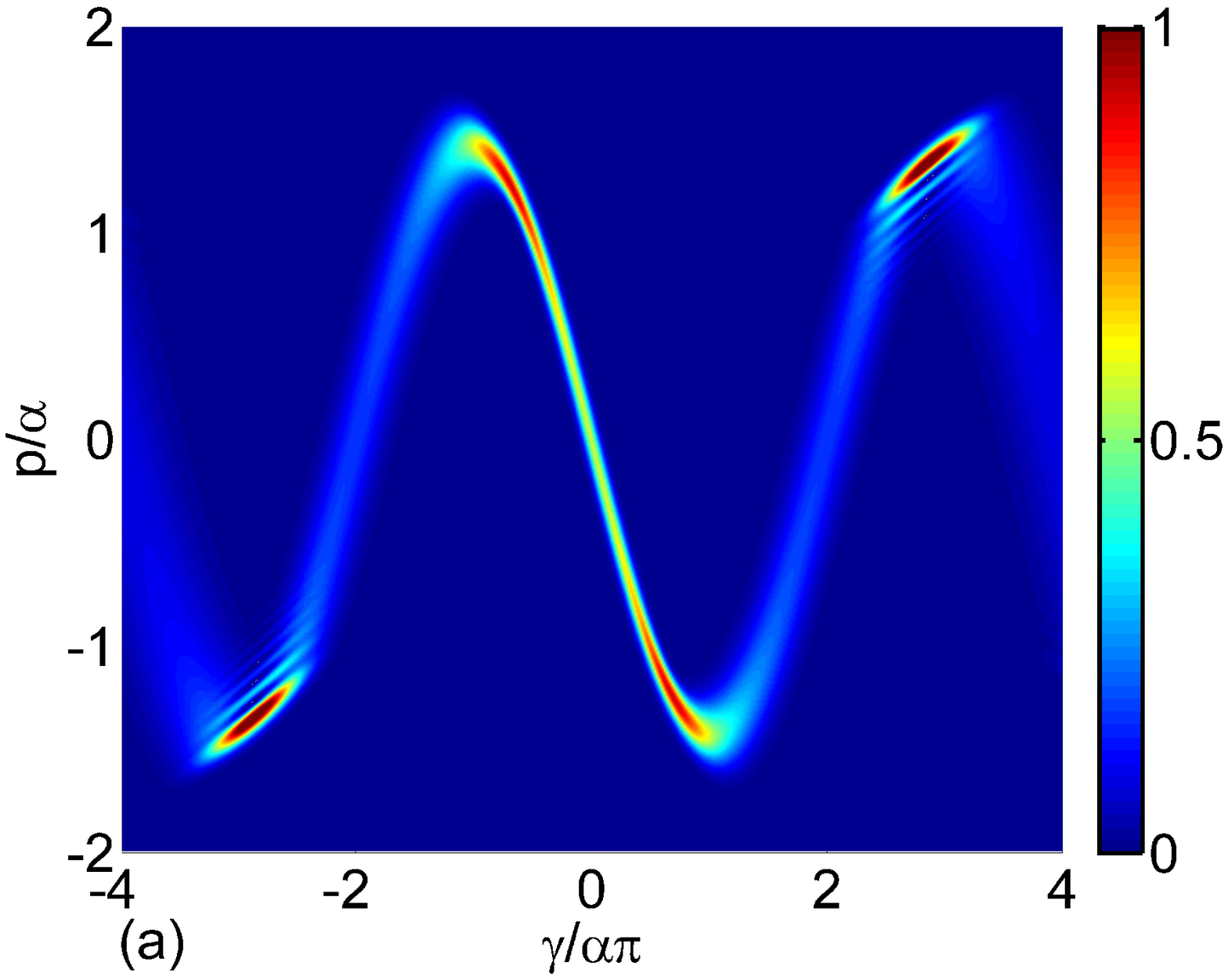} %
\includegraphics[bb=0 0 600 450, width=0.45\textwidth, clip]{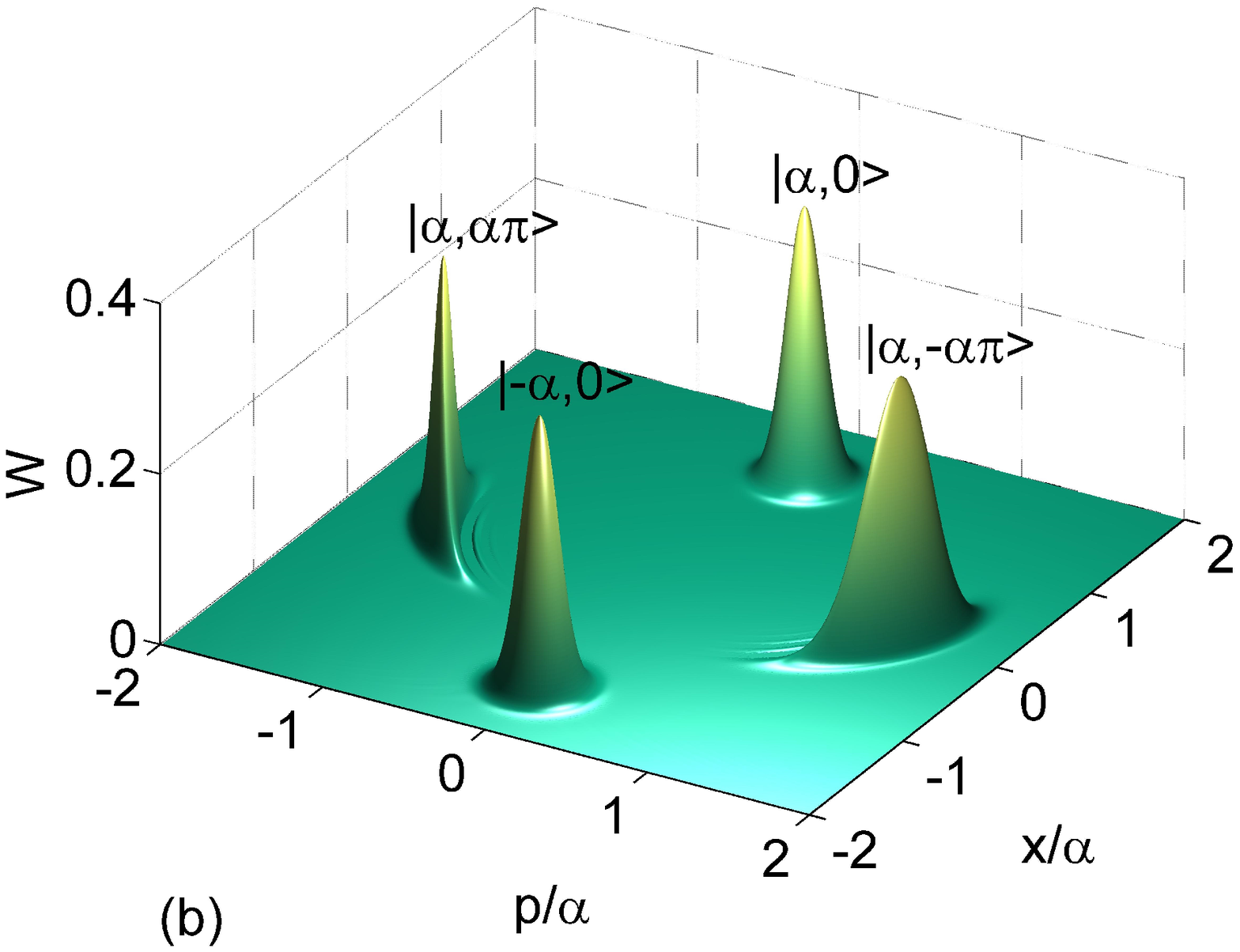}
\end{center}
\caption{(Color online) (a) Plot of the probability distribution $P\left( p,%
\protect\gamma \right) $\ defined in Eq. (\protect\ref{P_gamma} ) for $%
\protect\alpha =7$ and $\protect\gamma \in \left[ -4\protect\alpha\protect%
\pi ,4\protect\alpha\protect\pi \right] $. It shows that the states $%
\left\vert \protect\alpha ,\pm \protect\gamma \right\rangle $ are also
sufficiently distinguishable when $\protect\gamma $ is around the values of $%
\protect\alpha \protect\pi (2l+1)$, $l=0$, $\pm 1$, $\pm 2$, $...$. (b) Plot
of Wigner functions for states $\left\vert \pm \protect\alpha %
,0\right\rangle $, and $\left\vert \protect\alpha ,\pm \protect\gamma %
\right\rangle $ with $\protect\alpha =7$ and $\protect\gamma =\protect\alpha
\protect\pi $ . }
\label{figure1}
\end{figure}

By taking $\alpha =7$, $\gamma \in \left[ -4\alpha \pi ,4\alpha \pi \right] $%
, we plot the probability distribution%
\begin{equation}
P\left( p,\gamma \right) =\left\vert \psi _{\gamma }\left( p\right)
\right\vert ^{2}=e^{-\left\vert \alpha \right\vert ^{2}}\left\vert \sum_{n}%
\frac{\alpha ^{n}}{\sqrt{n!}}\mathrm{e}^{-\mathrm{i}\gamma \sqrt{n}}\phi
_{n}\left( p\right) \right\vert ^{2},  \label{P_gamma}
\end{equation}%
where $\phi _{n}\left( p\right) $\ is the eigenfunction of the harmonic
oscillator system in $p$ space.

We plot the probability distribution $W_{\alpha ,\gamma }(x,p)$ for $\alpha $
$=\pm 7$, $\gamma =0$; $\alpha =7$, $\gamma =\pm \alpha \pi $. Fig. \ref%
{figure1} shows that $\left\vert \alpha ,\pm \gamma \right\rangle $\ are
also sufficiently distinguishable when $\gamma $ is around the values of $%
\alpha \pi (2l+1)$, with $l=0$, $\pm 1$, $\pm 2$, $...$. It is not true for
the cases of $\left\vert l\right\vert \gg 1$, which is beyond our concern
because they cannot be precisely prepared as $\left\vert l\right\vert $\
increases in our scheme. It indicates that the state $c_{1}\left\vert \alpha
,\gamma \right\rangle +c_{2}\left\vert \alpha ,-\gamma \right\rangle $ can
be considered as a Schr\"{o}dinger cat state. Moreover, modified coherent
state $\left\vert \alpha ,\gamma \right\rangle $\ is physically relevant. It
can be prepared from the canonical coherent state $\left\vert \alpha
,0\right\rangle $\ by natural time evolution in a system with $\sqrt{n}$\
spectrum. In the Ref. \cite{JGea-Banacloche91},\ it is shown that $%
\left\vert \alpha ,\gamma \right\rangle $\ can be obtained from $\left\vert
\alpha ,0\right\rangle $\ by natural time evolution. However, we would like
to point that the approximation in the\ Ref. \cite{JGea-Banacloche91}\
\begin{eqnarray}
&&\sum_{n=0}^{\infty }\alpha ^{n}/\sqrt{n!}\mathrm{e}^{\mp \mathrm{i}\sqrt{n}%
\lambda t}  \nonumber \\
&\approx &\mathrm{e}^{\mp \mathrm{i}\alpha \lambda t/2}\sum_{n=0}^{\infty
}\alpha ^{n}/\sqrt{n!}\mathrm{e}^{\mp \mathrm{i}n\lambda t/2\alpha },
\end{eqnarray}%
is not suitable for the purpose of quantum information processing, since the
norm of the overlap between the exact and approximate wavefunctions is
estimated as%
\begin{eqnarray}
&&\left\vert \mathrm{e}^{-\left\vert \alpha \right\vert
^{2}}\sum_{n=0}^{\infty }\frac{\alpha ^{2n}}{n!}\mathrm{e}^{-\mathrm{i}\pi
\left( \alpha \sqrt{n}-n/2\right) }\right\vert  \nonumber \\
&\approx &\left\vert \frac{1}{\sqrt{2\pi }\alpha }\int_{-\infty }^{\infty
}\exp \left[ -\left( \frac{1}{2\alpha ^{2}}-\frac{\mathrm{i}\pi }{8\alpha
^{2}}\right) \left( x-\alpha ^{2}\right) ^{2}\right] \mathrm{d}x\right\vert
\nonumber \\
&=&0.8868,
\end{eqnarray}%
which is far from $1$. This is one of the reasons we are interested in the
MCS.\textbf{\ }In the following section, we will demonstrate that the
atom-field interaction can induce effective nonlinear spectrum of the
photon. The cat state can be prepared by mapping the qubit state onto the
field.

\section{Effective separation of atom and field}

\label{sec_Effective separation of atom and field}In this section, we
investigate the dynamics of the JC model and present the connection to the
MCS. It has been pointed in the Ref. \cite{JGea-Banacloche91}\ that when the
field is initially in a coherent state with large average photon number and
the atom state is $\frac{1}{\sqrt{2}}\left( e^{-i\varphi }\left\vert
e\right\rangle \pm \left\vert g\right\rangle \right) $, the time evolution
of the atom-field system is\textbf{\ }%
\begin{equation}
\frac{1}{\sqrt{2}}\left( \mathrm{e}^{-\mathrm{i}\varphi }\mathrm{e}^{\mp
\mathrm{i}\frac{\lambda t}{2\alpha }}\left\vert g\right\rangle \pm
\left\vert e\right\rangle \right) \mathrm{e}^{-\alpha ^{2}/2}\sum_{n}\frac{%
\alpha ^{n}}{\sqrt{n!}}\mathrm{e}^{-\mathrm{i}n\varphi }\mathrm{e}^{\mp
\mathrm{i}\sqrt{n}\lambda t}\left\vert n\right\rangle .
\end{equation}
It indicates that a MCS can be generated from a natural time evolution. To
reveal the underlying mechanism of this process, we will reconsider this
issue in an alternative way. The obtained result is applicable for a class
of field states and will shed the light on the scheme of quantum information
transfer from field to atom.

Consider a single-cavity JC model
\begin{eqnarray}
H &=&\lambda \left( \sigma _{+}a+\sigma _{-}a^{\dag }\right) +\frac{1}{2}%
\omega _{a}\sigma _{z}+\omega a^{\dagger }a,  \label{H_JC} \\
\sigma _{+} &=&\left( \sigma _{-}\right) ^{\dagger }=\left\vert
e\right\rangle \left\langle g\right\vert ,\sigma _{z}=\left\vert
e\right\rangle \left\langle e\right\vert -\left\vert g\right\rangle
\left\langle g\right\vert ,
\end{eqnarray}%
where $\omega $ is photon frequency, $\left\vert g\right\rangle $ and $%
\left\vert e\right\rangle $\ denote the ground and excited states of atom
with transition frequency $\omega _{a}$, and $\lambda $ is the atom-field
coupling constant. Under the resonance condition $\omega _{a}=\omega $, it
can be reduced to a simple form%
\begin{equation}
H=\lambda \left( \sigma _{+}a+\sigma a^{\dagger }\right) ,  \label{H_r}
\end{equation}%
in the interaction picture. We notice that the Jaynes-Cummings model has
been realized in the laboratory in several well-known ways \cite%
{Harochegroup,Winelandgroup,Kimblegroup,JLi13,J-MPirkkalainen13} and
employed to engineer states by using atom as a medium since almost two
decades \cite{CKLaw96}.

We note that the excitation number, $\mathcal{N}=\frac{1}{2}\sigma
_{z}+a^{\dagger }a+\frac{1}{2}$ is a conservative quantity, \textit{i.e.}, $%
\left[ \mathcal{N},H\right] =0$. So the Hamiltonian can be diagonalized in
each $2\times 2$ invariant subspace. We start our analysis from Hamiltonian (%
\ref{H_r}), which can be rewritten in the form%
\begin{equation}
H=H_{\mathrm{1}}+H_{\mathrm{2}},
\end{equation}%
\begin{eqnarray}
H_{\mathrm{1}} &=&\lambda (\mathrm{e}^{\mathrm{i}\varphi }\left\vert
e\right\rangle \left\langle g\right\vert \sum_{n}\sqrt{n+1}\left\vert
n\right\rangle \left\langle n\right\vert +\mathrm{e}^{-\mathrm{i}\varphi
}\left\vert g\right\rangle \left\langle e\right\vert \sum_{n}\sqrt{n}%
\left\vert n\right\rangle \left\langle n\right\vert ,) \\
H_{\mathrm{2}} &=&\lambda \sum_{n}\sqrt{n+1}\left( \mathrm{e}^{\mathrm{i}%
\varphi }\left\vert e\right\rangle \left\langle g\right\vert \left\vert
n\right\rangle -\left\vert g\right\rangle \left\langle e\right\vert
\left\vert n+1\right\rangle \right) \left( \mathrm{e}^{-\mathrm{i}\varphi
}\left\langle n+1\right\vert -\left\langle n\right\vert \right) ,
\end{eqnarray}%
by taking $\left\langle n+1\right\vert =$ $\mathrm{e}^{\mathrm{i}\varphi
}\left( \left\langle n\right\vert +\mathrm{e}^{-\mathrm{i}\varphi
}\left\langle n+1\right\vert -\left\langle n\right\vert \right) $, $%
\left\langle n\right\vert $ $=\mathrm{e}^{-\mathrm{i}\varphi }\left\langle
n+1\right\vert $ $-\mathrm{e}^{-\mathrm{i}\varphi }\left\langle
n+1\right\vert +\left\langle n\right\vert $, where $\varphi $\ is an
arbitrary real number.

For a separated state
\begin{equation}
\left\vert \phi \left( 0\right) \right\rangle =\left( c_{1}\left\vert
g\right\rangle +c_{2}\left\vert e\right\rangle \right)
\sum_{n}f_{n}\left\vert n\right\rangle ,  \label{Phi_0}
\end{equation}%
where $\left\vert c_{1}\right\vert ^{2}+\left\vert c_{2}\right\vert ^{2}=1$%
,\ and $\sum_{n}\left\vert f_{n}\right\vert ^{2}=1$, one can reduce the
Hamiltonian to the simple form under certain conditions.\ We note that if
\begin{equation}
f_{n}\approx \mathrm{e}^{\mathrm{i}\varphi }f_{n-1}\text{, }\left\vert
f_{n}\right\vert ^{2}\ll 1,  \label{app cond_1}
\end{equation}%
we have $\left\vert H_{\mathrm{2}}\left\vert \phi \left( 0\right)
\right\rangle \right\vert \ll \left\vert H_{\mathrm{1}}\left\vert \phi
\left( 0\right) \right\rangle \right\vert $, \textit{i.e.}, the dynamics of
state $\left\vert \phi \left( 0\right) \right\rangle $\ is governed by the
Hamiltonian $H_{\mathrm{1}}$\ approximately. Now we focus on the Hamiltonian
$H_{\mathrm{1}}$. In addition to the condition in Eq. (\ref{app cond_1}),
when we consider the field state satisfying%
\begin{equation}
\bar{n}\gg \triangle n\gg 1,  \label{cond2}
\end{equation}%
where $\bar{n}$ denotes the average photon number with $\bar{n}$ $%
=\sum_{n_{1}}^{n_{2}}n\left\vert f_{n}\right\vert ^{2}$, $\triangle
n=n_{2}-n_{1}$, we can have $H_{\mathrm{1}}=H_{+}P_{+}+H_{-}P_{-}$. Here the
projection operators $P_{\pm }$ for atom state are $P_{\pm }=\frac{1}{2}%
\left( \left\vert g\right\rangle \pm \mathrm{e}^{\mathrm{i}\varphi
}\left\vert e\right\rangle \right) \left( \left\langle g\right\vert \pm
\mathrm{e}^{-\mathrm{i}\varphi }\left\langle e\right\vert \right) $, which
satisfy $P_{+}+P_{-}=1$ and $P_{+}P_{-}=0$. In the derivation, we have used
the approximation $\left( n+1\right) ^{1/2}\approx n^{1/2}+\frac{1}{2\bar{n}%
^{1/2}}$, under the condition in Eq. (\ref{cond2}). The sub-Hamiltonians $%
H_{\pm }$ are in the form $H_{+}=-H_{-}=\frac{\lambda }{2\bar{n}^{1/2}}%
\left\vert e\right\rangle \left\langle e\right\vert +\lambda \sum_{n}\sqrt{n}%
\left\vert n\right\rangle \left\langle n\right\vert $.

Before further discussion of the implication of the obtained result, two
distinguishing features need to be mentioned. First, the spectra of the
photons in $H_{\pm }$ are $\pm \lambda \sqrt{n}$, which are related to the
preparation of states $\left\vert \alpha ,\pm \gamma \right\rangle $ from $%
\left\vert \alpha ,0\right\rangle $ by natural time evolution, as mentioned
above. This is crucial for the scheme to write the qubit state in the field.
Second, we note that $H_{+}$ and $H_{-}$ have opposite sign, which leads to
the time evolution of photons in two reverse directions respectively. This
is crucial for the scheme to read the state from the field.

\section{Quantum state transfer between atom and field}

\begin{figure}[tbp]
\begin{center}
\includegraphics[bb=12 440 582 829,width=0.7\textwidth,
clip]{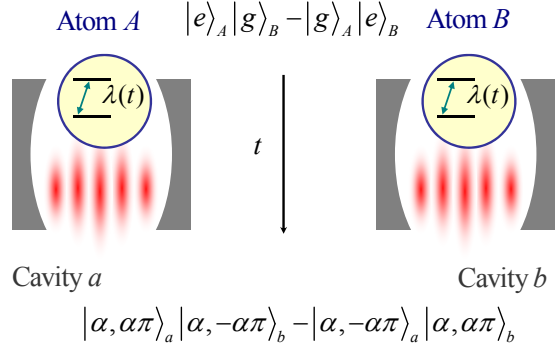}
\end{center}
\caption{(Color online) Schematic illustration of the system for the
entanglement transfer protocol. Atoms $A$ and $B$ are embedded in their
respective cavities $a$ and $b$. Initially, two atoms are maximally
entangled.\ The natural time evolution driven by the $A-a$ and $B-b$
interactions $\protect\lambda \left( t\right) $ can transfer the $AB$
entanglement to the fields entanglement between cavities $a$ and $b$. The
time dependence of coupling constant $\protect\lambda \left( t\right) $\ is
explained in the text.}
\label{cavity}
\end{figure}

\label{sec_Quantum state transfer between atom and field}A state in the form
of Eq. (\ref{Phi_0}) can be rewritten as $\left\vert \phi \left( 0\right)
\right\rangle =\frac{1}{\sqrt{2}}\left[ A\left( \left\vert g\right\rangle +%
\mathrm{e}^{\mathrm{i}\varphi }\left\vert e\right\rangle \right) +B\left(
\left\vert g\right\rangle -\mathrm{e}^{\mathrm{i}\varphi }\left\vert
e\right\rangle \right) \right] \times \sum_{n}\left\vert f_{n}\right\vert
\mathrm{e}^{\mathrm{i}n\varphi }\left\vert n\right\rangle $, where $A=\frac{1%
}{\sqrt{2}}\left( c_{1}+\mathrm{e}^{-\mathrm{i}\varphi }c_{2}\right) $ and $%
B $ $=\frac{1}{\sqrt{2}}\left( c_{1}-\mathrm{e}^{-\mathrm{i}\varphi
}c_{2}\right) $. Then we have%
\begin{eqnarray}
\left\vert \phi \left( t\right) \right\rangle = &&\frac{1}{\sqrt{2}}A\left(
\left\vert g\right\rangle +\mathrm{e}^{-\mathrm{i}\frac{\lambda t}{2\bar{n}%
^{1/2}}}\mathrm{e}^{\mathrm{i}\varphi }\left\vert e\right\rangle \right)
\sum_{n}\left\vert f_{n}\right\vert \mathrm{e}^{-\mathrm{i}\sqrt{n}\lambda t}%
\mathrm{e}^{\mathrm{i}n\varphi }\left\vert n\right\rangle  \nonumber \\
&&+\frac{1}{\sqrt{2}}B\left( \left\vert g\right\rangle -\mathrm{e}^{\mathrm{i%
}\frac{\lambda t}{2\bar{n}^{1/2}}}\mathrm{e}^{\mathrm{i}\varphi }\left\vert
e\right\rangle \right) \sum_{n}\left\vert f_{n}\right\vert \mathrm{e}^{%
\mathrm{i}\sqrt{n}\lambda t}\mathrm{e}^{\mathrm{i}n\varphi }\left\vert
n\right\rangle .  \label{phi_t}
\end{eqnarray}%
It is the superposition state of two independent evolution processes in
which there are no interactions between atom and field. At the instants, $%
t_{l}=\left( 2l+1\right) \tau $, $l=0,1,2,...$, $\tau =\bar{n}^{1/2}\pi
/\lambda $, we have
\begin{equation}
\left\vert \phi \left( t_{l}\right) \right\rangle =\frac{\left\vert
g\right\rangle -\left( -1\right) ^{l}\mathrm{i}\mathrm{e}^{\mathrm{i}\varphi
}\left\vert e\right\rangle }{\sqrt{2}}\left( A\left\vert \Phi
_{+}^{l}\right\rangle +B\left\vert \Phi _{-}^{l}\right\rangle \right) ,
\label{t_l}
\end{equation}%
where $\left\vert \Phi _{\pm }^{l}\right\rangle =\sum_{n}\left\vert
f_{n}\right\vert \mathrm{e}^{\mp \mathrm{i}\sqrt{n}\lambda t_{l}}\mathrm{e}^{%
\mathrm{i}n\varphi }\left\vert n\right\rangle $. As one can see in the
formula above, an arbitrary atom state can retrieve a pure state at the
instants $t_{l}$. Remarkably, the initial atomic state $\left( A,B\right) $
is mapped on the field state if $\left\vert \Phi _{+}^{l}\right\rangle $ and
$\left\vert \Phi _{-}^{l}\right\rangle $\ are orthogonal, while the atom is
always in the state $\left( \left\vert g\right\rangle -\mathrm{i}\mathrm{e}^{%
\mathrm{i}\varphi }\left\vert e\right\rangle \right) /\sqrt{2}$. This is
termed as \textquotedblleft attractor\textquotedblright\ in Ref. \cite%
{JGea-Banacloche91}, which considered the initial field state being a
coherent state. However, according to our analysis, the initial state can be
a class of field states. To demonstrate this point, we consider the simplest
field state, a top-hat state as the form%
\begin{equation}
f_{n}=\left\{
\begin{array}{c}
1/\Delta \text{, }\bar{n}-\Delta /2\leqslant n<\bar{n}+\Delta /2 \\
0,\text{elsewhere}%
\end{array}%
\right. .
\end{equation}%
Submitting $f_{n}$\ to Eq. (\ref{phi_t}) and taking $\varphi =0$, $t=\tau =%
\bar{n}^{1/2}\pi /\lambda $, yields%
\begin{eqnarray}
\left\vert \phi \left( \tau \right) \right\rangle = &&\frac{1}{\sqrt{2}}%
A\left( \left\vert g\right\rangle -\mathrm{i}\left\vert e\right\rangle
\right) \frac{1}{\Delta }\sum_{n-\Delta /2}^{n+\Delta /2-1}\mathrm{e}^{-%
\mathrm{i}\sqrt{n}\bar{n}^{1/2}\pi }\left\vert n\right\rangle  \nonumber \\
&&+\frac{1}{\sqrt{2}}B\left( \left\vert g\right\rangle -\mathrm{i}\left\vert
e\right\rangle \right) \frac{1}{\Delta }\sum_{n-\Delta /2}^{n+\Delta /2-1}%
\mathrm{e}^{\mathrm{i}\sqrt{n}\bar{n}^{1/2}\pi }\left\vert n\right\rangle .
\end{eqnarray}%
To verify this approximate result, we compute the norm of the overlap $%
\digamma \left( t\right) $\ between the analytical and numerical final
states. For the cases of $\Delta =5$, $10$, and $20$, $\bar{n}=49$, we have $%
\digamma \left( \tau \right) =$\ $0.9001$, $0.9955$, and $0.9738$. It
indicates that for a given $\bar{n}$, an optimal $\Delta $\ can lead to a
perfect fidelity.\ Thus the coherent state $\left\vert \alpha \right\rangle $%
\ is not the unique field state leading to the separation of the atom and
field.

Nevertheless, we still take the coherent state $\left\vert \alpha
\right\rangle $\ as an example to illustrate our scheme in this paper. From
Eq. (\ref{t_l}), we note that the initial state $\left( A,B\right) $\ can
never go back to the atom as expected. This procedure can be employed to
transfer or store the quantum information to the field. However, the stored
information can not be read out from the field via the further time
evolution alone this path, as expected in a general scheme, the initial
state is revival periodically.

Qubit decoherence times are on the order of a few microseconds, in
particular, that for the transmon is about $\sim $4 $\mu $s. As a result,
the implementation, including the preparation of the cat state and adiabatic
adjustment of the parameters, should be accomplished within the time much
shorter than the decoherence time. According to recent experiments, the
decoherence time can be a few dozen times longer than the $\tau $\cite%
{OEM11,Liu13}.

Considering the initial field state as a coherent state $\left\vert \alpha
,0\right\rangle $, with $f_{n}=\mathrm{e}^{-\left\vert \alpha \right\vert
^{2}/2}\alpha ^{n}/\sqrt{n!}$ and $\varphi =0$, we have
\begin{equation}
\left\vert \phi \left( \tau \right) \right\rangle =\frac{1}{\sqrt{2}}\left(
\left\vert g\right\rangle -\mathrm{i}\left\vert e\right\rangle \right)
\left( A\left\vert \alpha ,\alpha \pi \right\rangle +B\left\vert \alpha
,-\alpha \pi \right\rangle \right) \text{.}
\end{equation}%
Then the quantum information in the atom is encoded into the field. We would
like to point that, the initial state of the atom cannot be revival again as
expected, not like the swap gate.

Another interesting feature in such dynamic process is that two effective
Hamiltonians for atoms differ only in an opposite sign, \textit{i.e.}, $%
H_{+}=-H_{-}$. It shows that two atom states $\left( \left\vert
g\right\rangle +\mathrm{e}^{\mathrm{i}\varphi }\left\vert e\right\rangle
\right) /\sqrt{2}$ and $\left( \left\vert g\right\rangle -\mathrm{e}^{%
\mathrm{i}\varphi }\left\vert e\right\rangle \right) /\sqrt{2}$ evolve in
the same way but in opposite time directions. It is crucial for the scheme
of reading out the information from the field.

A trivial way to retrieve the initial atomic state is to switch the sign of
interaction strength $\lambda $\ to realize the reversed time evolution.
Alternative operation can also implement the same task: flipping the states $%
\left( \left\vert g\right\rangle -\mathrm{i}\mathrm{e}^{\mathrm{i}\varphi
}\left\vert e\right\rangle \right) /\sqrt{2}\rightarrow \left( \left\vert
g\right\rangle +\mathrm{i}\mathrm{e}^{\mathrm{i}\varphi }\left\vert
e\right\rangle \right) /\sqrt{2}$ by the external pulse. (rotation operator $%
\mathrm{e}^{\mathrm{i}\pi \left( \sigma _{z}+1\right) /2}=-\sigma _{z}$)
Then taking the state $\left\vert \widetilde{\phi }\left( 0\right)
\right\rangle =\left( \left\vert g\right\rangle +\mathrm{i}\left\vert
e\right\rangle \right) \left( A\left\vert \alpha ,\alpha \pi \right\rangle
+B\left\vert \alpha ,-\alpha \pi \right\rangle \right) /\sqrt{2}$, as the
initial state, we have $\left\vert \widetilde{\phi }\left( \tau \right)
\right\rangle =-\sigma _{z}\left\vert \phi \left( 0\right) \right\rangle $.

Although this conclusion is achieved in the framework of approximation, we
would like to point that such a time-reversal process is exact. The
underlying mechanism is similar to the phenomenon of spin echo. We note that
the Hamiltonian (\ref{H_r}) obeys the anti-commutation relation%
\begin{equation}
\left\{ \sigma _{z},H\right\} =0.  \label{A_C}
\end{equation}%
Then for an arbitrary initial state $\psi \left( 0\right) $ and an arbitrary
time interval\ $t$, we have $\psi ^{\prime }\left( 2t\right) $ $\equiv
\left( -\sigma _{z}\right) \mathrm{e}^{-\mathrm{i}Ht}\left( -\sigma
_{z}\right) \mathrm{e}^{-\mathrm{i}Ht}\psi \left( 0\right) $ $=\mathrm{e}^{%
\mathrm{i}Ht}\mathrm{e}^{-\mathrm{i}Ht}\psi \left( 0\right) =\psi \left(
0\right) $.

It represents a process similar to the spin echo, refocusing the information
spreading to the field. The atom retrieves\ its initial state under the
operation of spin flip at half time. We note that the relation in Eq. (\ref%
{A_C}) requires the resonance condition in an atom-field system. Then
resonance is crucial for the reversal process. This feature can be applied
to the atom-field system without rotating-wave-approximation. On the other
hand, the original Hamiltonian $H_{\text{\textrm{S}}}=H+\omega \left(
\mathcal{N-}\frac{1}{2}\right) $ in the Schrodinger picture has $\left\{
\sigma _{z},H_{\text{\textrm{S}}}\right\} \neq 0$. However, we still have $%
\psi _{\mathrm{s}}^{\prime }\left( 2t\right) \equiv \left( -\sigma
_{z}\right) e^{-\mathrm{i}H_{\mathrm{S}}t}\left( -\sigma _{z}\right) e^{-%
\mathrm{i}H_{\mathrm{S}}t}\psi \left( 0\right) $ $=\left( -\sigma
_{z}\right) e^{-\mathrm{i}Ht}e^{-\mathrm{i}\omega \left( \mathcal{N-}\frac{1%
}{2}\right) t}\left( -\sigma _{z}\right) e^{-\mathrm{i}Ht}e^{-\mathrm{i}%
\omega \left( \mathcal{N-}\frac{1}{2}\right) t}\psi \left( 0\right) $ $=e^{-2%
\mathrm{i}\omega \left( \mathcal{N-}\frac{1}{2}\right) t}e^{\mathrm{i}Ht}e^{-%
\mathrm{i}Ht}\psi \left( 0\right) $ $=e^{-2\mathrm{i}\omega \left( \mathcal{%
N-}\frac{1}{2}\right) t}\psi \left( 0\right) $, due to the relation $\left[
\sigma _{z},\mathcal{N}\right] =0$. We note that operator $e^{-2\mathrm{i}%
\omega \left( \mathcal{N-}\frac{1}{2}\right) t}$\ cannot induce any
unexpected effects for the initial state $\psi \left( 0\right) $\ we
consider in this paper.

\begin{figure}[tbp]
\begin{center}
\includegraphics[bb=-2 195 598 646,width=0.7\textwidth, clip]{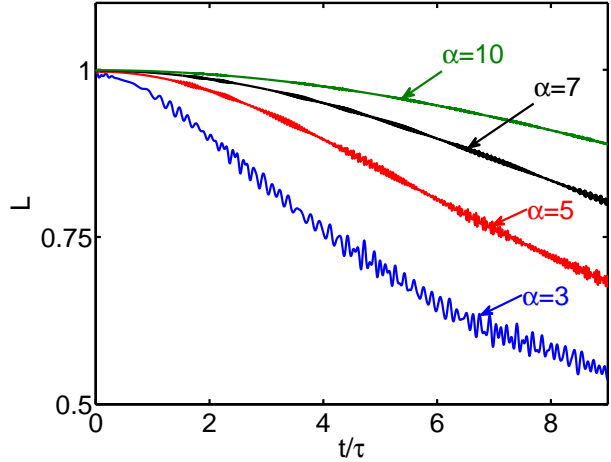}
\end{center}
\caption{(Color online) Plots of the Loschmidt echo $L\left( t\right) $
defined in Eq. (\protect\ref{L} ) for the cases of $\protect\alpha =3$, $5$,
$7$ and $10$. It indicates that the decay of $L\left( t\right) $\ becomes
slow as the average photon number increase.}
\label{figure4}
\end{figure}

We would like to point that the separation of atom and field is approximate.
In fact, the atom and field always interact with each other. The effective
separation is the result of interference. We note that the approximation
condition require that the phase between $\left\vert n\right\rangle $ and $%
\left\vert n+1\right\rangle $ is arbitrary but identical, \textit{i.e.}, $%
\varphi $\ is $n$-dependent. However, the evolved state will acquire an
extra phase being proportional to $\sqrt{n}t$\ rather than $nt$. Then, as
time increases, the deviation of the evolved wave function from the
approximation condition get large. In order to estimate the time scale
within which the effective Hamiltonian is available, we plot of the
Loschmidt echo%
\begin{equation}
L\left( t\right) =\left\vert \left\langle \psi \left( 0\right) \right\vert
\mathrm{e}^{\mathrm{i}Ht}\mathrm{e}^{-\mathrm{i}H_{\mathrm{1}}t}\left\vert
\psi \left( 0\right) \right\rangle \right\vert ,  \label{L}
\end{equation}%
which is a estimate of the differentiating effects for $H$\ and $H_{\mathrm{1%
}}$. From Fig. \ref{figure4}, we see that $L\left( t\right) $\ depends on
the $\alpha $ in the initial state. For $\alpha =10$, $L\left( t\right) $
can be more than $0.95$ at $t=\tau $, and $0.9$ within $t=8\tau $. Then we
see that the effective Hamiltonian is available in a short time.

In order to quantitatively evaluate the extent of approximation of the
effective Hamiltonian and demonstrate the write and read scheme, the
numerical method is employed to simulate the dynamic processes of quantum
state transfer.

For the writing process, we take the initial state as $\left\vert \phi
\left( 0\right) \right\rangle =\frac{1}{\sqrt{2}}\left( \left\vert
g\right\rangle +\left\vert e\right\rangle \right) \left\vert \alpha
,0\right\rangle $. The fidelity of the state transfer is $F_{W}\left(
t\right) =\left\vert \left\langle \phi \left( 0\right) \right\vert \mathrm{e}%
^{\mathrm{i}Ht}\left\vert \phi _{W}\right\rangle \right\vert $, where $%
\left\vert \phi _{W}\right\rangle =\left( \left\vert g\right\rangle -\mathrm{%
i}\left\vert e\right\rangle \right) /2\left\vert \alpha ,\alpha \pi
\right\rangle $\ is the target state. Similarly, the fidelity for the
reading process is defined as $F_{R}\left( t\right) =\left\vert \left\langle
\psi \left( 0\right) \right\vert \mathrm{e}^{\mathrm{i}Ht}\left\vert \psi
_{R}\right\rangle \right\vert ,$ where the initial state $\left\vert \psi
\left( 0\right) \right\rangle =-\sigma _{z}\left\vert \phi _{W}\right\rangle
$\ and the target state is $\left\vert \psi _{R}\right\rangle $ $=-\sigma
_{z}\left\vert \phi \left( 0\right) \right\rangle =\frac{1}{\sqrt{2}}\left(
\left\vert g\right\rangle -\left\vert e\right\rangle \right) \left\vert
\alpha ,0\right\rangle $. Straightforward derivation shows that $F_{W}\left(
t\right) =F_{R}\left( t\right) \equiv F\left( t\right) $. In Fig. \ref%
{figure2}, $F\left( t\right) $ is plotted for the cases of $\alpha =3,5,7,$
and $10$, which shows that the fidelity increases with $\alpha $ and the QST
approaches to perfect when the average photon number is more than two dozen.

\begin{figure}[tbp]
\begin{center}
\includegraphics[bb=10 200 560 625,width=0.45\textwidth, clip]{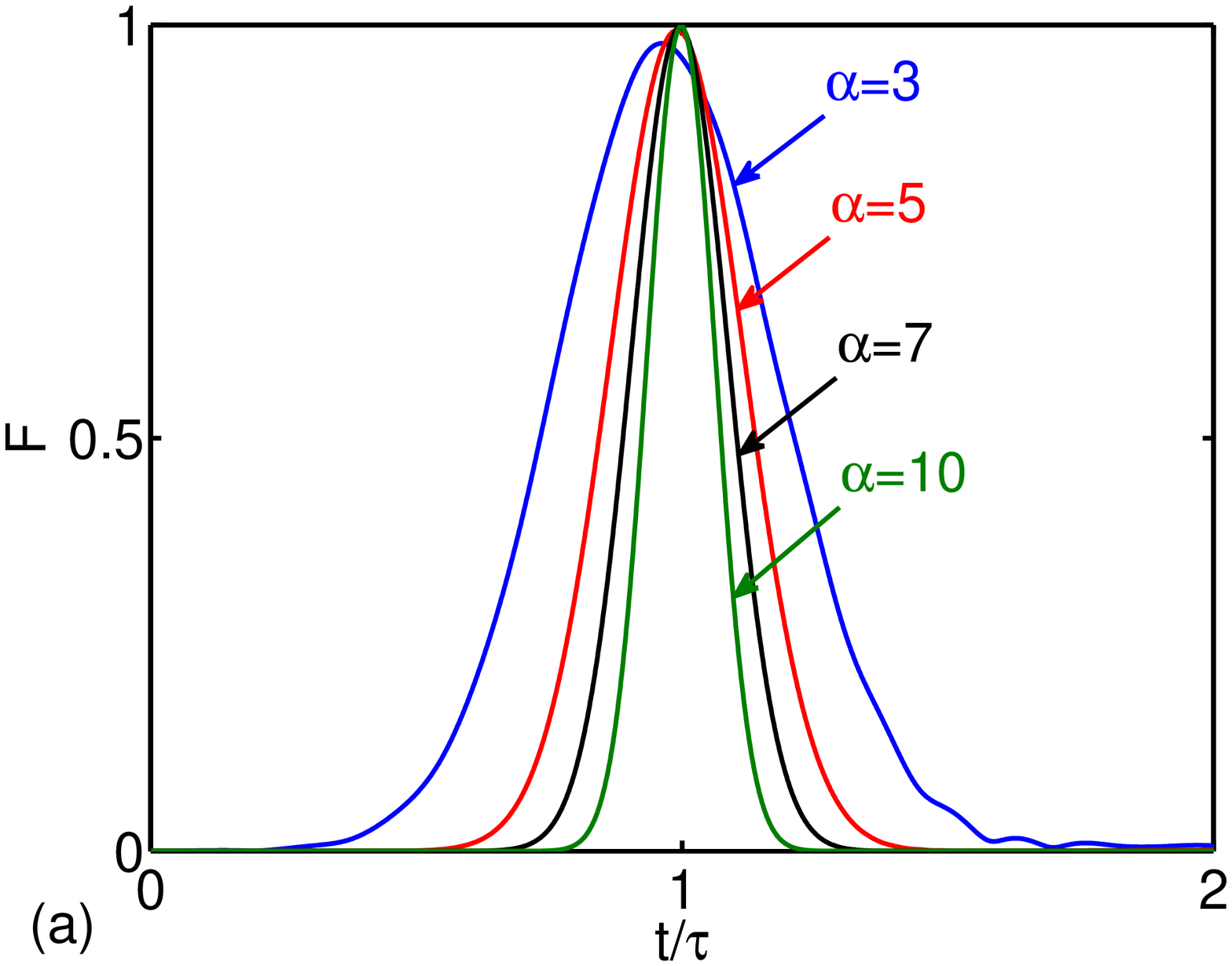} %
\includegraphics[bb=10 200 560 625,width=0.45\textwidth, clip]{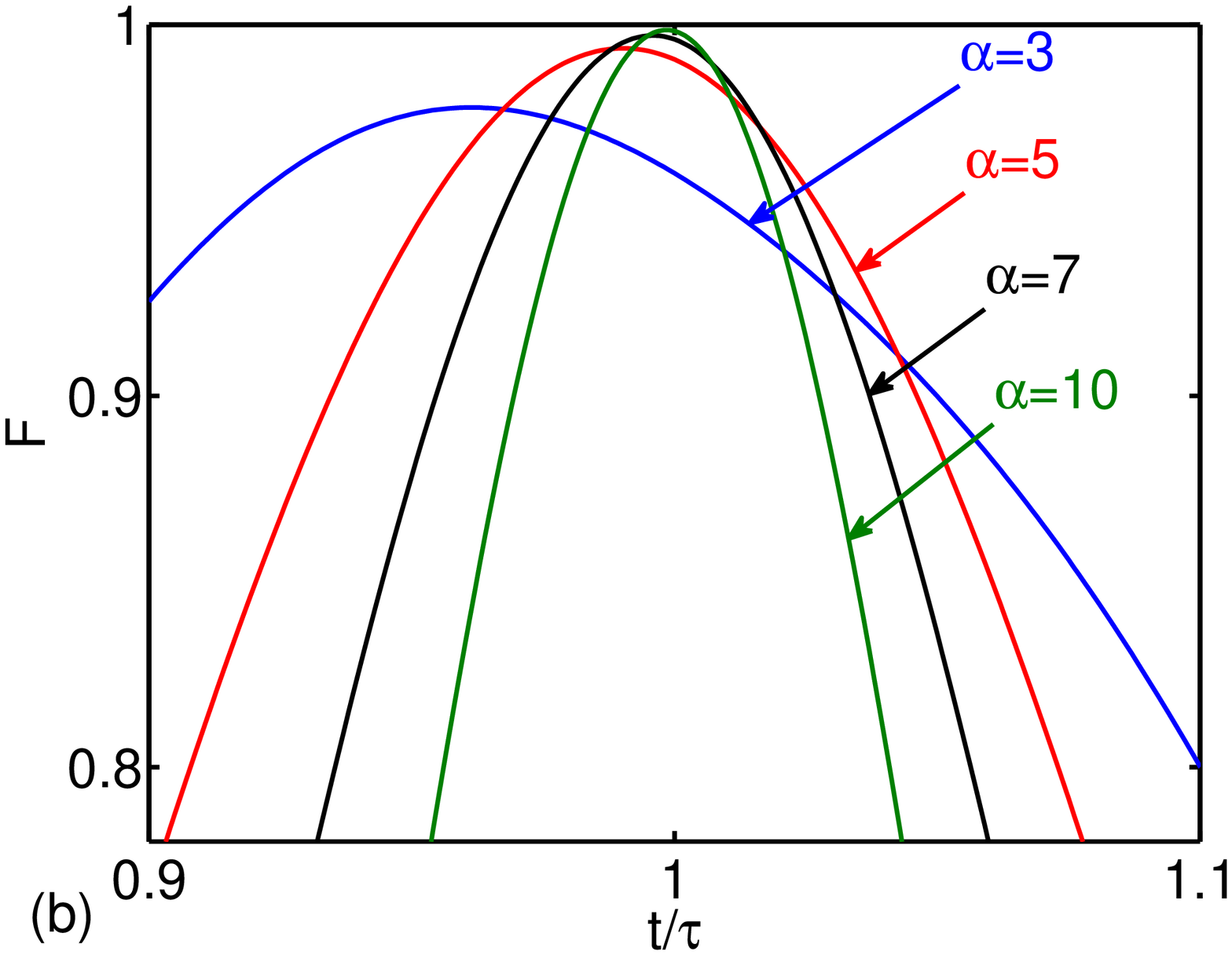}
\end{center}
\caption{(Color online) Plots of the fidelity $F\left( t\right) $\ as
function of time for the cases of $\protect\alpha =3$, $5$, $7$ and $10$.
(b) is a zoom-in figure of (a). The fidelity $F\left( t\right) $\ in each
cases has a maximum at the instant $t_{m}$. Both\ $F\left( t_{m}\right) $
and $t_{m}/\protect\tau $\ tend to $1$ as $\protect\alpha $ increases.}
\label{figure2}
\end{figure}

\section{Entanglement transfer}

\label{sec_Entanglement transfer}A straightforward application of the above
analysis is generation of a non-local Schr\"{o}dinger cat state, which is a
fundamental resource in fault-tolerant quantum computing and quantum
communication. In this section, we study the dynamics of entanglement
transfer in a system composed of two initially entangled atoms, each located
in one of two non-interacting cavities. A schematic illustration of the
system is given in Fig. \ref{cavity}. The Hamiltonian of the set-up is \cite%
{J-PGazeau09}

\begin{figure}[tbp]
\begin{center}
\includegraphics[bb=0 200 550 625,width=0.45\textwidth, clip]{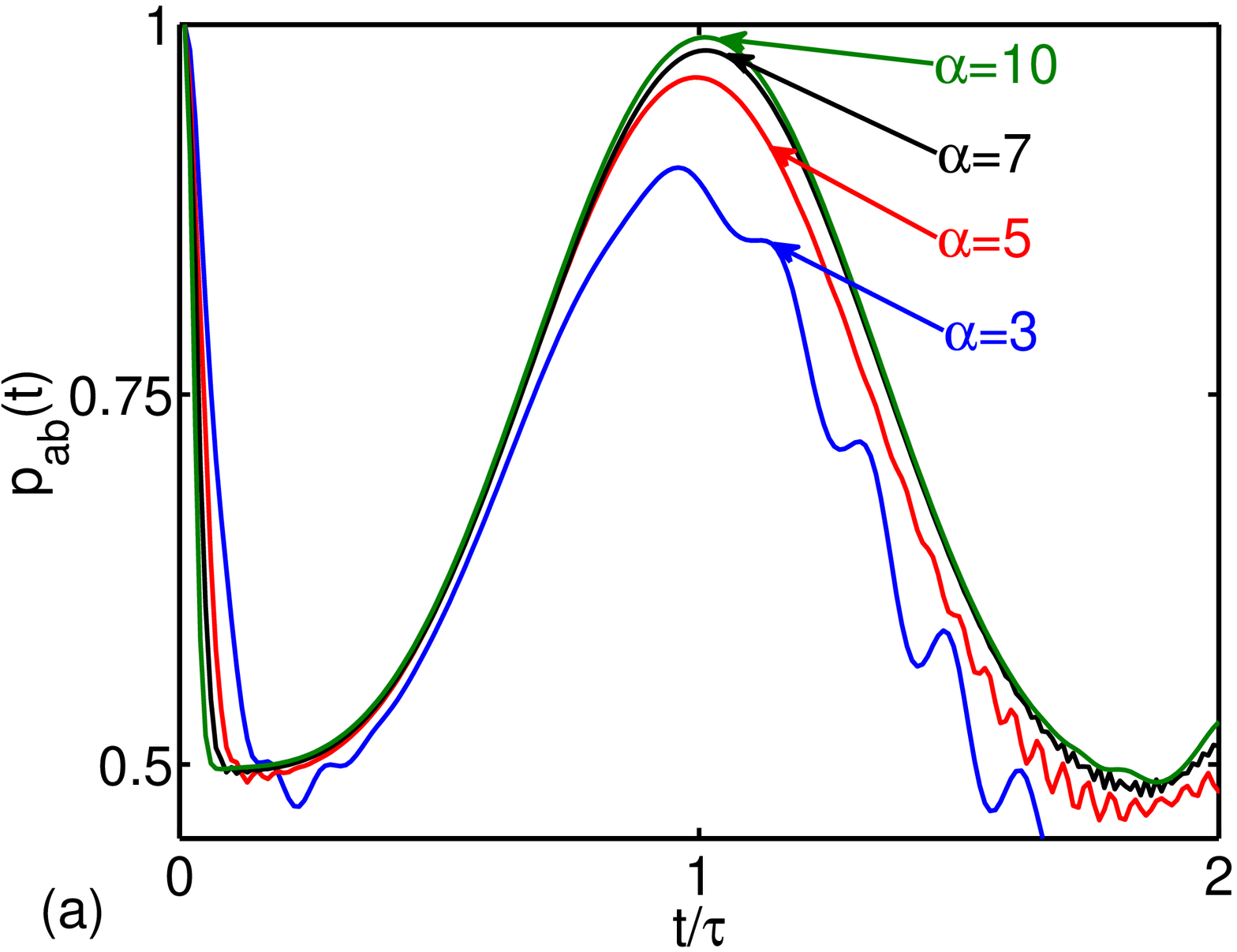} %
\includegraphics[bb=0 204 550 629,width=0.45\textwidth, clip]{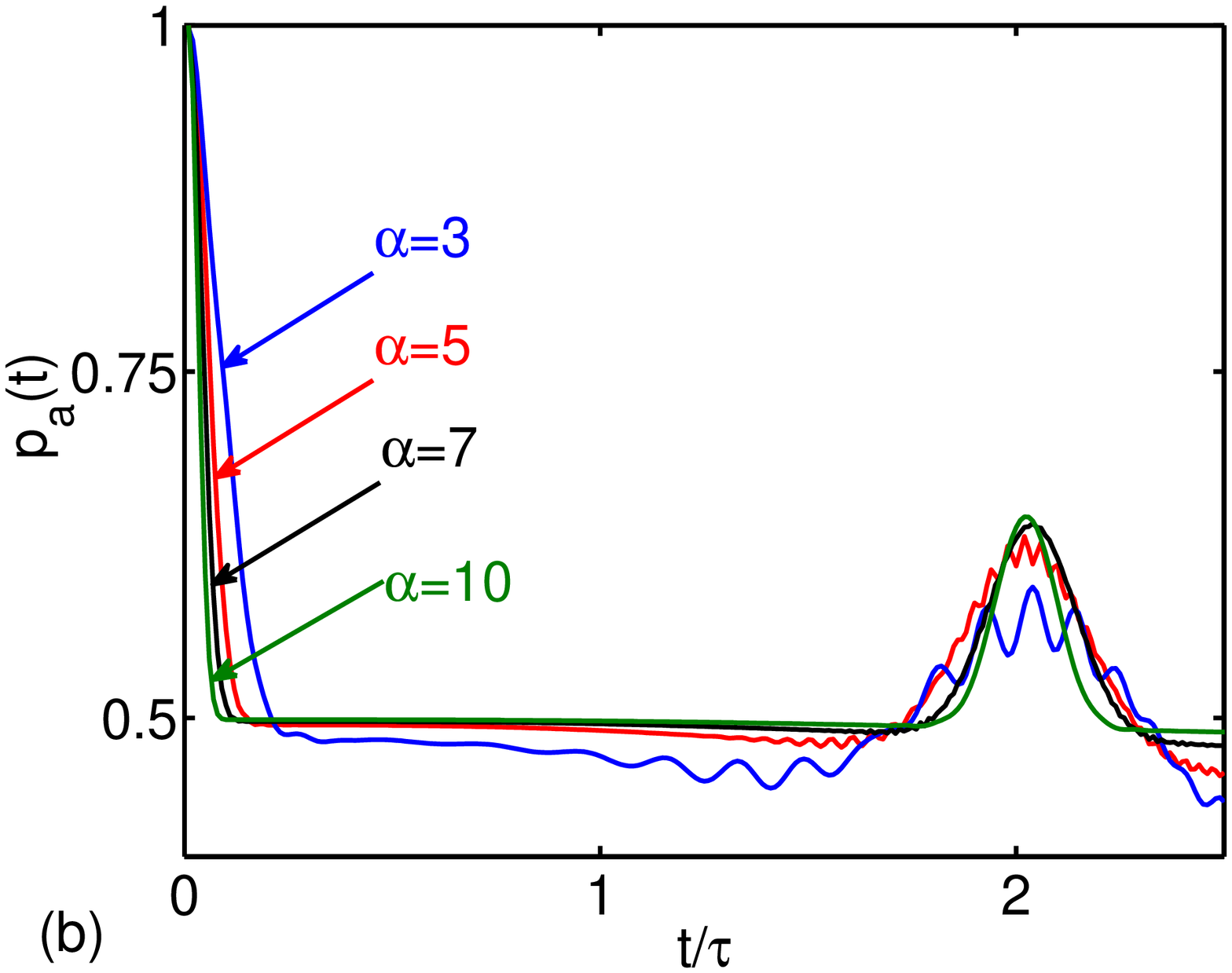}
\end{center}
\caption{(Color online) Plot of the purities $p_{ab}\left( t\right) $ and $%
p_{a}\left( t\right) $ for the type of initial state in Eq. (\protect\ref%
{initial AB}) as a function of time (in unit of $\protect\tau =\protect%
\alpha \protect\pi /\protect\lambda $) and parameter $\protect\alpha $. It
shows that the maximal entanglement between two distant cavities can be
generated for large $\protect\alpha $.}
\label{figure3}
\end{figure}

\begin{equation}
H_{AB}=\lambda (a^{\dagger }\sigma _{-}^{A}+b^{\dagger }\sigma _{-}^{B}+%
\mathrm{H.c.}),  \label{H_AB}
\end{equation}%
where $\sigma _{-}^{A}(\sigma _{-}^{B})$ and $a^{\dagger }(b^{\dagger })$
are the corresponding operators of the atom and field in the cavity $A(B)$,
respectively, and $\lambda $ is the coupling constant between the atoms and
their cavities. Here we only consider the case of resonance.

The separation of systems $A$ and $B$ makes the dynamics of whole system
easier to be understood \cite{MYonac07}. It is predictable that the
entanglement of two atoms can be perfectly transferred to the fields $%
a^{\dagger }$ and $b^{\dagger }$, which leads to generate a non-local Schr%
\"{o}dinger cat state in large $\alpha $\ limit.

For finite $\alpha $, to demonstrate the process of entanglement transfer,
we compute the purity $p_{ab}$ of the field states in two cavities, as well
as that $p_{a}$\ (or $p_{b}$) for a single cavity. The former is the measure
of the entanglement between the atoms and the fields, while the later is
that between fields in two cavities. Here the purities for an arbitrary
state $\left\vert \phi \right\rangle $ are defined as $p_{ab}=\mathrm{Tr}%
\left( \rho _{ab}\right) ^{2}$, $p_{a}=\mathrm{Tr}\left( \rho _{a}\right)
^{2},$ with $\rho _{ab}=\mathrm{Tr}_{\mathrm{AB}}\left( \left\vert \phi
\right\rangle \left\langle \phi \right\vert \right) $, $\rho _{a}=\mathrm{Tr}%
_{\mathrm{b}}\left( \rho _{ab}\right) $, where \textrm{Tr}$_{\mathrm{AB}}(.)$%
\ and \textrm{Tr}$_{\mathrm{b}}(.)$\ denote the operation of tracing out all
atomic states and field $b$ states, respectively.

Now we examine the efficiency of the entanglement transfer from the atoms to
the field for finite $\alpha $. The initial state is
\begin{equation}
\left\vert \phi \left( 0\right) \right\rangle =\frac{\left\vert
e\right\rangle _{A}\left\vert g\right\rangle _{B}-\left\vert g\right\rangle
_{A}\left\vert e\right\rangle _{B}}{\sqrt{2}}\left\vert \alpha
,0\right\rangle _{a}\left\vert \alpha ,0\right\rangle _{b},
\label{initial AB}
\end{equation}%
which denotes a maximally entangled $AB$\ state, but an unentangled $ab$\
state. According to our analysis above mentioned, for sufficient large $%
\alpha $, the state evolves to
\begin{eqnarray}
\left\vert \phi \left( \tau \right) \right\rangle &=&\frac{1}{2\sqrt{2}}%
\left( \left\vert g\right\rangle _{A}-\mathrm{i}\left\vert e\right\rangle
_{A}\right) \left( \left\vert g\right\rangle _{B}-\mathrm{i}\left\vert
e\right\rangle _{B}\right)  \nonumber \\
&&\left( \left\vert \alpha ,\alpha \pi \right\rangle _{a}\left\vert \alpha
,-\alpha \pi \right\rangle _{b}-\left\vert \alpha ,-\alpha \pi \right\rangle
_{a}\left\vert \alpha ,\alpha \pi \right\rangle _{b}\right) ,
\end{eqnarray}%
at the instant $\tau $. We can see that the atoms $AB$\ state is
unentangled. Furthermore, fields $a$\ and $b$ are maximally entangled\ if
each cavity is\ regarded as a qubit. Quantities $p_{ab}$\ and $p_{a}$\ (or $%
p_{b}$) of the state $\left\vert \phi \left( \tau \right) \right\rangle $\
reflect this feature: Purity $p_{ab}=1$ indicates that the field state is in
pure state, being separable from the state of atoms. Purity $p_{a}=0.5$\
indicates the maximal entanglement fields $a$\ and $b$ as a two-qubit
system. In this sense, the later is only an essential condition\ for the
existence of state $\left\vert \phi \left( \tau \right) \right\rangle $.
Nevertheless, we believe that the simultaneous occurrence of $p_{ab}=1$\ and
$p_{a}=0.5$\ can be regarded as the evidence of the existence of the state $%
\left\vert \phi \left( \tau \right) \right\rangle $\ in the context of the
present model.

The purities $p_{ab}\left( t\right) $\ and $p_{a}\left( t\right) $\ for the
state evolving from the initial state given by\ Eq. (\ref{initial AB})\ are
plotted in Fig. \ref{figure3}\ for the cases of $\alpha =3$, $5$, $7$ and $%
10 $. As seen from the plots, purities $p_{ab}\left( t\right) $\ and $%
p_{a}\left( t\right) $\ approach to $1$\ and $0.5$,\ respectively,\ at time $%
\tau $\ as $\alpha $\ increases. In order to estimate the time scales
associated with the initial decay, we employ the curve fitting for the
numerical results in Fig. \ref{figure3} and obtained the fitted function as $%
p_{a}\left( t\right) \approx p_{b}\left( t\right) $ $\approx 0.5\exp \left(
-\lambda ^{2}t^{2}\right) +0.5$ $=0.5\exp \left[ -\alpha ^{2}\pi ^{2}\left(
t/\tau \right) ^{2}\right] +0.5$. By the same procedure, the fitted function
around the revival can be obtained as $p_{ab}\left( t\right) $ $\approx
0.5\sin ^{4}\left( \lambda t/2\alpha \right) \exp \left( -\lambda
^{2}t^{2}/8\alpha ^{4}\right) +0.5$ $=0.5\sin ^{4}\left[ \pi \left( t/\tau
\right) /2\right] \exp \left[ -\pi ^{2}\left( t/\tau \right) ^{2}/8\alpha
^{2}\right] +0.5$, which indicates the width of the revival as $w\approx
0.73\tau $ $=0.73\alpha \pi /\lambda $. It shows that the entanglement can
be perfectly transferred from atoms to fields. It also provides an
alternative scheme for preparation of non-local superpositions of
quasi-classical light states.

Before ending this paper, we want to stress that there is an important
feature of the scenario. Actually, the dynamic processes in above schemes
are invariant if the coupling strength $\lambda $\ is time dependent. All
the derivation above is still true if we replace $\lambda \Delta t$ by $%
\int_{0}^{\Delta t}\lambda \left( t\right) dt$. One can take the interaction
in the form $\lambda \left( t\right) =\lambda _{0}\mathrm{exp}\left( -\alpha
^{2}t^{2}\right) $\ which can be used to turn an initial coherent field into
a Schr\"{o}dinger cat state. It ensures\ the switching control feasible for
schemes in experiment. In addition, the swiching processes of $\lambda $\ in
two cavities cannot be simultaneous. This alllows the generation of
entanglement between a cavity and a distant atom.

Nevertheless, there is always dissipation in cavities. The effect of
dissipation on a macroscopic superposition of quantum states has been
studied with the use of a Markovian master-equation approach. An approximate
solution was given for the JC model with cavity losses by J.Gea-Banacloche
in 1993 \cite{JGea-Banacloche93}. It has been reported that, although the
dissipation destroys the coherence of the macroscopic superposition very
rapidly, preparation and observation of the cat state should be possible,
which depends on a factor $F\left( t\right) $\ characterizing the cavity
quality, or describing the effect of dissipation. According to their
analysis, when\textbf{\ }the magnitude of $F\left( t\right) $ is close to $1$%
, the effect of dissipation is slight. More explicitly, at the revival time $%
t_{R}=2\tau $, the corresonding factor yields%
\begin{equation}
\left\vert F\left( t_{R}\right) \right\vert \approx e^{-4\pi \kappa \bar{n}%
^{3/2}/\lambda },  \label{F}
\end{equation}%
where $\kappa $ is the damping rate of photon to the reservoir at zero
temperature. In our paper, because there is no interaction between two
cavities, the effect of dissipation should be the same as the result in ref.
\cite{JGea-Banacloche93}. So the outcome should be observable in the system
with $\kappa /\lambda \sim 10^{-5}$, which predicts $\left\vert F\left(
t_{R}\right) \right\vert \approx 0.9844$, when $\bar{n}=25$.

However, according to the current experiment, the realization of our scheme
is difficult due to the dissipation. It is reported that parameters $\lambda
=2\pi \times 75\mathrm{MHz}$ and $\kappa =2\pi \times 3.5\mathrm{MHz}$ can
be achievable in optical cavities with the wave-length in the region $%
630-850 $\textrm{nm} in recent experiments \cite{Spillane05,Buck03}, which
result in $\left\vert F\left( t_{R}\right) \right\vert \approx 0$. In
another case, the optical fiber decay at a $852$nm wavelength is about $2.2$%
\textrm{dB/km} \cite{Gordon04,Zheng10}, which corresponds to the fiber decay
rate of $\kappa =0.152\mathrm{MHz}$. In this case, the $\left\vert F\left(
t_{R}\right) \right\vert \approx 0.6025$, which is still lower.

\section{Summary}

\label{sec_Summary}

In summary, we have presented a scheme for state transfer between atom and
cavity field in Jaynes-Cummings model. It is shown that the nonlinearity
arising from the atom-field coupling can induce the generation of modified
coherent states, which can be shown to be macroscopically distinguishable as
standard coherent states. We have shown that an arbitrary atom state can be
mapped onto the field as Schr\"{o}dinger cat state via natural time
evolution. The reversal process can also be achieved in the aid of
spin-echo-like technique. We also found that the coherent state is just one
of a class of field states, which can be used as an initial state to perform
this task. This result can be extended to non-interacting multi-cavity
system. Analytical and numerical calculations have demonstrated that the
dynamic process on two-cavity system can provide an alternative scheme for
preparation of non-local superpositions of quasi-classical light states.
Moreover, we discussed the effect of dissipation.

\section*{Acknowledgments}

We acknowledge the support of the National Basic Research Program (973
Program) of China under Grant No. 2012CB921900 and CNSF (Grant No. 11374163).

\end{document}